\documentclass[letterpaper]{article} 
\usepackage{aaai2026}  
\usepackage{amsfonts} 
\usepackage{times}  
\usepackage{helvet}  
\usepackage{courier}  
\usepackage[hyphens]{url}  
\usepackage{graphicx} 
\usepackage{natbib}  
\usepackage{caption} 
\usepackage{algorithm}
\usepackage{algpseudocode}

\usepackage{newfloat}
\usepackage{listings}
\DeclareCaptionStyle{ruled}{labelfont=normalfont,labelsep=colon,strut=off} 
\floatstyle{ruled}
\newfloat{listing}{tb}{lst}{}
\floatname{listing}{Listing}
\nocopyright

\title{Domain Design for the Cops and Robbers Problem}
\author{
    Connor Little,
    Meagan Mann,
    Erin Meger,
    Christian Muise
}
\affiliations{%
$\lbrace$ connor.little, 20mmm27, erin.meger, christian.muise $\rbrace$ @queensu.ca \\
  Queen's University\\
  Kingston ON, Canada
}

\begin{document}

\maketitle

\begin{abstract}
  Cops and Robbers is a well-studied problem in graph theory. The setting consists of a robber and one or more cops placed on an undirected graph. Taking turns moving throughout the graph, the cops try to capture the robber. The property of interest is whether $k$ cops suffice to ensure at least one cop occupies the same vertex as the robber, after a finite number of turns, given any configuration of their initial placement; if successful, the graph is referred to as ``$k$-copwin''. In this work, we cast the problem of determining whether a graph is $k$-copwin as a non-deterministic planning problem and use state-of-the-art planners to compute this property. The cop movement is cast as non-deterministic movement (to capture all possible strategies), while the robber movement is deterministic in nature. We also extend the base model using several variations from the graph theory literature. 
  %
  Our work demonstrates that planning can efficiently solve the problem, opening the door to systematic analysis of this property on arbitrary graphs.
\end{abstract}

\section{Introduction}
\label{sec:introduction}
The Cops and Robbers problem is a popular topic in graph theory \cite{chudnovsky2024cops, enright2023cops, pralat2016meyniel}. It simulates a chase between a robber and a specified number of cops across a graph. The robber tries to avoid capture forever, while the cops want to occupy the same vertex as the robber. If $k$ cops are able to successfully capture the robber, then we call the graph $k$-copwin. The cop number $c(G)$ of the graph is the smallest $k$. Despite its simple rules, a myriad of research has focused on both techniques for solving the game and variations on extending it \cite{drunkrobber2012, frieze2010variationscopsrobbers, CLARKE2025394, bonato2017characterizationsalgorithmsgeneralizedcops}.

Cops and Robbers is  prototypical EXPTIME problem in Graph Theory \cite{kinnersley2015cops}, with its most elusive property being Meyneil's Conjecture\cite{frankl1987cops}: \[c(G) = O\left( \sqrt{|V(G)|}\right)\]

This problem is of high interest and where other current strategies are lacking, using planning in this area can be a powerful tool. To translate the problem into planning, we introduce a new perspective: the non-deterministic cop. This method of modelling the problem has each cop make a non-deterministic move each turn. As such, the planner must create policies for all reachable states, corresponding to a policy for all cop strategies.
By coupling this with a neutral initial state that allows the robber and the cops to choose their starting configurations, we can search for a policy that considers every configuration on a single graph simultaneously. This allows the model to exhaustively test and explore the environment. Results of this type have previously not been tackled by researchers in this area, and most improvements in research rely on underlying graph structure \cite{kenter2025improved} or asymptotic analyses \cite{scott2011bound, lu2012meyniel}.

More precisely, we translate the problem into a fully observable non-deterministic (FOND) planning problem. Under these rules, all state information is known, and there is no hidden information. This is typical in Cops and Robbers research, as there is little to be gained by limiting what information the cops and the robbers have access to, since any significant change to this would be considered a new variant.
The cop's actions are non-deterministic: when moving, an arbitrary node is chosen. Non-deterministic actions are not guaranteed to be fair, but this lack of fairness does not impact the correctness of our result.

Due to certain limitations, FOND models cannot be compared in terms of optimality. Instead, we will take a subjective approach to how these policies are used: we will compare them based on interpretability. This subjectivity will provide insights for theoretical directions. The benefit of automated planning is that it allows us to efficiently create policies for the robber.  If a policy exists that allows the robber to evade capture indefinitely, we can conclude that the graph is not $k$-cop win.

In addition to creating the problems and domains for the standard version, we explore some variations of the Cops and Robbers problem. We have implemented two versions: the variable-move-speed robber \cite{Frieze2012} and the friendly robber. The variable-move-speed robber can take multiple moves per turn. This allows the robber to cover more ground and avoid capture where they otherwise could not. 
The friendly robber's objective is to be captured as soon as possible when it begins on a random starting node not occupied by a cop. This inverts the problem to one of being caught by a cop rather than avoiding them. The former variation has been studied extensively, while the latter offers a new perspective on the problem.

Finally, we empirically test our solution by applying it to classes of graphs and finding the cop number for the entire class. We use these generated sets to determine what (if any) graph properties can imply the existence of a $k$-copwin strategy. Our work shows promising results for using automated planning to solve and explore properties of entire classes of complex problems.

\section{Background}
\label{sec:background}

\subsection{The Cops and Robbers Problem}
The game of Cops and Robbers is a variation of Pursuit-Evasion played on a graph. It was first introduced by Nowakowski and Winkler \cite{Nowakowski1983} in 1983, and independently by Quilliot \cite{QUILLIOT198655}. It is played  with perfect information between two adversarial forces: the $k \geq 1$ cops and the single robber. Starting with the robber, the players take turns making moves across a connected, undirected, simple graph. Each player takes turns moving via a single edge to a new vertex. The robber aims to avoid capture indefinitely, while the cop aims to occupy the same vertex as the robber, called capture. 

The interesting problem with this game is to determine the cop number of a given graph, $G$. That is, to determine if $G$ is $k$-copwin for some $k$.  We formalize the Cops and Robbers problem specifically by giving a graph $G$ and a constant $k \in \mathbb{Z}$ and determining if $G$ is or is not $k$-copwin. We iterate over increasing values of $k$ until success, thereby obtaining the cop number of the graph.

A simple strategy for the cops is to put a cop on every single node of the graph, guaranteeing capture instantly. This implies an upper bound on the cop number is $|V(G)|$. The lower bound is similarly straightforward, as we know we must have at least one cop to capture. However, determining $c(G)$ exactly is often quite challenging, and in fact for general graphs is EXPTIME-complete \cite{kinnersley2015cops}.  

\subsection{Variations on the Cops and Problems}
Frieze et al.introduced many of the popular variations on the problem \cite{Frieze2012}. The first variation covered is the directed graph. In this variation, the graph has one-way edges, resulting in a much more restricted search space. The second variation is the fast robber. With this condition, the robber may move a variable number of spaces each turn, with the maximum distance predetermined. This allows the robber much more maneuverability when avoiding the cops. This can be extended into the infinitely fast robber, who may jump to any unguarded node. 

Another more recent variant is the drunk-robber variant first introduced by Komarov and Winkler \cite{komarov2013}. In this variation, the robber employs no strategy; instead, it wanders aimlessly throughout the graph. This has been expanded to the ``tipsy cop and drunk robber'' in which the robber still moves aimlessly and the cop may make knowledgeable moves or may move randomly \cite{Harris2020}. There is also the case of the invisible robber \cite{Kehagias2013}. This variant has the robber's position unknown until capture. While these variants are similar, none fully exemplify the non-deterministic cop that we use in this paper. 

\subsection{Automated Planning and Evasion Problems}
In automated planning, work on Cops and Robbers and related games is extremely limited. Even outside the domain of automated planning, using AI to investigate the Cops and Robbers problem has not been explored in depth. To our knowledge, there have been no such attempts for Cops and Robbers, but there has been progress in adjacent areas of pursuit-evasion. Nussbaum and Yörükçü studied the problem of Moving Target Search (MTS), in which an agent attempts to capture another non-stationary agent \cite{nussbaum}. The key inclusion in their algorithm is the use of environment abstraction to reduce the search space without increasing the algorithm's cost. As such, they were able to improve upon all compared algorithms in all tested environments. 

Planning under visibility constraints is also a common occurrence in the pursuit-evasion planning literature. 
Guibas et al. introduce both an algorithm for solving the problem and bounds on necessary conditions for capturing the evader \cite{guibas1999visibility}. Their paper uses a polygonal environment instead of a graph. Stiffler and Kane look at the domain of search and rescue \cite{9341031}. The evaders are victims who must be caught (rescued) and may move unpredictably. A key feature is that the algorithm must guarantee that each evader is found. They use a forward-search algorithm and an algorithm that simulates evader trajectories to aid in path computation.

More recently, Macindoe et al. studied a variant of the Cops and Robbers problem within an automated planning framework \cite{macindoe2023}. In their work, they substantially modified the original game, in a way that is distinct from this implementation. 

\section{A Planning Model for Cops and Robbers}
\label{sec:approach}

The model we present is for the non-deterministic Cops and Robbers problem. No variations are included in this model unless explicitly mentioned. The standard version of cops and robbers starts with a graph $G=\{V, E\}$, where $V$ is a set of vertices and $E$ is a set of edges. 
The $k$ cops are placed on vertices of $V \backslash \{v\}$. In each round, the robber makes a move along one incident edge to some adjacent vertex, and then all $k$ cops make a similarly defined move. This is done until the robber is unable to traverse an edge without being on an occupied node, or until the depth limit is reached.

We use standard terminology from automated planning in the nondeterministic formalism. In automated planning, predicates are properties or relationships of objects that can be either true or false given a state. When predicates are instantiated with specific objects, they become facts that define the state. An action consists of preconditions that must hold for it to be able to be executed and effects describing how the state changes. Nondeterministic effects are expressed with a \textbf{oneof} clause, meaning exactly one of several possible outcomes will occur. 

In nondeterministic planning, a strong solution guarantees that the goal will be reached under all possible circumstances. A strong cyclic solution allows cycles in the states, and still guarantees that the goal will be reached under fairness assumption. A strong acyclic solution is a strong solution without loops.

To begin, we define the following predicates:

\begin{itemize}
    \item (at ?x - entity ?y - location)
    \begin{itemize}
        \item This fluent is true if the entity occupies said location.
    \end{itemize}
    \item (edge ?x ?y - location)
    \begin{itemize}
        \item This fluent defines the architecture of the graph.
    \end{itemize}
    \item (turn ?x - cop)
    \begin{itemize}
        \item This fluent determines which entities may make a move. It is the robber's turn if all turns are set to false.
    \end{itemize}

    \item (caught)
    \begin{itemize}
        \item This fluent is true if the robber is ever caught.
    \end{itemize}
    \item (survived)
    \begin{itemize}
        \item This fluent determines if the robber is still alive.
    \end{itemize}
    \item (done)
    \begin{itemize}
        \item This fluent is a flag for ending the game.
    \end{itemize}
    \item (nil)
    \begin{itemize}
        \item This fluent does nothing. It is a placeholder for the identity action.
    \end{itemize}
    \item (move0), (move1), etc
    \begin{itemize}
        \item These fluent counts the number of moves a robber may make.
    \end{itemize}
\end{itemize}

\begin{algorithm}
\caption{Robber Movement}\label{alg:1}
\begin{algorithmic}[1]
\State (:action robber\_move
\State \hspace{\algorithmicindent} :parameters (?x ?y - location)
\State \hspace{\algorithmicindent} :precondition (and
\State \hspace{\algorithmicindent}\hspace{\algorithmicindent} (at rob1 ?x)
\State \hspace{\algorithmicindent}\hspace{\algorithmicindent} (edge ?x ?y)
\State \hspace{\algorithmicindent}\hspace{\algorithmicindent} (forall (?c - cop) (not (turn ?c))))
\State \hspace{\algorithmicindent} :effect (and
\State \hspace{\algorithmicindent}\hspace{\algorithmicindent} (when (exists (?c - cop) (at ?c ?y)) (caught)))
\State \hspace{\algorithmicindent}\hspace{\algorithmicindent} (not (at rob1 ?x))
\State \hspace{\algorithmicindent}\hspace{\algorithmicindent} (at rob1 ?y)
\State \hspace{\algorithmicindent}\hspace{\algorithmicindent}
\State \hspace{\algorithmicindent}\hspace{\algorithmicindent} (when (move0) 
\State \hspace{\algorithmicindent}\hspace{\algorithmicindent} (and (not move0) (move1)))
\State \hspace{\algorithmicindent}\hspace{\algorithmicindent}
\State \hspace{\algorithmicindent}\hspace{\algorithmicindent} ...
\State \hspace{\algorithmicindent}\hspace{\algorithmicindent} 
\State \hspace{\algorithmicindent}\hspace{\algorithmicindent} (when (moveK) 
\State \hspace{\algorithmicindent}\hspace{\algorithmicindent}  (and (not moveK) (forall (?c - cop)  (turn ?c))))
\State \hspace{\algorithmicindent}\hspace{\algorithmicindent}
\State \hspace{\algorithmicindent}\hspace{\algorithmicindent} (oneof (survived) (and))
\State \hspace{\algorithmicindent} )
\State )
\end{algorithmic}
\end{algorithm}

Next, we may define some of the important actions. Algorithm \ref{alg:1} defines the movement of the robber. Since there is only 1 robber, the only parameter this action requires is the location. Line 4 ensures the robber is at the first location. Line 5 ensures there is a connection between nodes. Line 6 ensures it is not the cop's turn. The effect is as follows. Line 8 checks for a collision between the robber and the cop to determine the game's end. Lines 9 and 10 remove the robber from the first node and place it in the second node. Line 11 increments the robber's move counter. On the last move, on line 12, all cops have their turn set to be true. Lastly, line 13 non-deterministically ends the game. This is necessary as we do not want the game to end once the robber is caught. The objective of the algorithm is to generate a plan for the robber to avoid for as long as possible. By choosing this method, the robber is incentivized to continuously avoid the cop without directly declaring a utility or reward function.

To add the variation of multiple moves, an additional set of statements is added. Each move is given a fluent: move0, move1, ..., moven. The first time a robber makes a move, move0 will be the only fluent set to true. For each move, a line is called that increments the move fluents. The first iteration will set move0 to false and move1 to true, the second will set move1 to false and move2 to true, and so on. Only on the last call will the cops' turns be set to true. In addition, each cop must now set move0 to true. This adds some redundant computation, but, as we demonstrate in the results, it is still an efficient implementation of this variation. 

\begin{algorithm}
\caption{Cop Movement Unoptimized}\label{alg:2}
\begin{algorithmic}[1]
\State (:action cop\_move
\State \hspace{\algorithmicindent} :parameters (?c - cop ?x ?y ?z - location)
\State \hspace{\algorithmicindent} :precondition (and
\State \hspace{\algorithmicindent}\hspace{\algorithmicindent} (turn ?c)
\State \hspace{\algorithmicindent}\hspace{\algorithmicindent} (num\_edges ?x n2)
\State \hspace{\algorithmicindent}\hspace{\algorithmicindent} (at ?c ?x)
\State \hspace{\algorithmicindent}\hspace{\algorithmicindent} (edge ?x ?y)
\State \hspace{\algorithmicindent}\hspace{\algorithmicindent} (edge ?x ?z)
\State \hspace{\algorithmicindent}\hspace{\algorithmicindent} (not (= ?y ?z)))
\State \hspace{\algorithmicindent} :effect (and
\State \hspace{\algorithmicindent}\hspace{\algorithmicindent} (not (turn ?c))
\State \hspace{\algorithmicindent}\hspace{\algorithmicindent} (move0)
\State \hspace{\algorithmicindent}\hspace{\algorithmicindent} (not (at ?c ?x))
\State \hspace{\algorithmicindent}\hspace{\algorithmicindent} (oneof 
\State \hspace{\algorithmicindent}\hspace{\algorithmicindent}\State \hspace{\algorithmicindent}\hspace{\algorithmicindent} (and (at ?c ?y) (when (at rob1 ?y) (caught)) )
\State \hspace{\algorithmicindent}\hspace{\algorithmicindent}\State \hspace{\algorithmicindent}\hspace{\algorithmicindent} (and (at ?c ?z) (when (at rob1 ?z) (caught)) )
\State \hspace{\algorithmicindent}\hspace{\algorithmicindent} )
\State \hspace{\algorithmicindent} )
\State )
\end{algorithmic}
\end{algorithm}

\begin{algorithm}
\caption{Cop Movement}\label{alg:3}
\begin{algorithmic}[1]
\State (:action cop\_move\_node\_2
\State \hspace{\algorithmicindent} :parameters (?c - cop)
\State \hspace{\algorithmicindent} :precondition (and
\State \hspace{\algorithmicindent}\hspace{\algorithmicindent} (turn ?c)
\State \hspace{\algorithmicindent}\hspace{\algorithmicindent} (at ?c node2))
\State \hspace{\algorithmicindent} :effect (and
\State \hspace{\algorithmicindent}\hspace{\algorithmicindent} (not (turn ?c))
\State \hspace{\algorithmicindent}\hspace{\algorithmicindent} (move0)
\State \hspace{\algorithmicindent}\hspace{\algorithmicindent} (not (at ?c node2))
\State \hspace{\algorithmicindent}\hspace{\algorithmicindent} (oneof 
\State \hspace{\algorithmicindent}\hspace{\algorithmicindent}\State \hspace{\algorithmicindent}\hspace{\algorithmicindent} (and (at ?c node1) 
\State \hspace{\algorithmicindent}\hspace{\algorithmicindent} (when (at rob1 node1) (caught)) )
\State \hspace{\algorithmicindent}\hspace{\algorithmicindent}\State  \hspace{\algorithmicindent}\hspace{\algorithmicindent} (and (at ?c node3) 
\State \hspace{\algorithmicindent}\hspace{\algorithmicindent} (when (at rob1 node3) (caught)) )
\State \hspace{\algorithmicindent}\hspace{\algorithmicindent} )
\State \hspace{\algorithmicindent} )
\State )
\end{algorithmic}
\end{algorithm}

Algorithm \ref{alg:2} defines the movement of the cop. The number of nodes adjacent to a given node is bounded. We need to numerically quantify the degree of the node before problem creation. The number of edge preconditions will grow linearly with the number of adjacent nodes. The number of inequality preconditions will grow quadratically with respect to the triangle numbers. The number of branches in the \textbf{oneof} clause will also grow linearly. To avoid this growth, we use the programmatically generated version for each graph. This creates an action for every node that a cop may take, by giving each problem a custom domain. Algorithm \ref{alg:3} shows an example of node2 with connections to node1 and node3. Both versions have some slight redundancy: they always set move0 to true. This is required to restart the robber's move cycle.

\begin{algorithm}
\caption{Terminate-Game}\label{alg:4}
\begin{algorithmic}[1]
\State (:action termination
\State \hspace{\algorithmicindent} :parameters ()
\State \hspace{\algorithmicindent} :precondition (and
\State \hspace{\algorithmicindent}\hspace{\algorithmicindent} (survived)
\State \hspace{\algorithmicindent}\hspace{\algorithmicindent} (not (caught)))
\State \hspace{\algorithmicindent} :effect (done)
\State )
\end{algorithmic}
\end{algorithm}

In addition to these actions, some supplementary actions are provided. The terminate-game action requires the variables 'survived' and not 'caught', returning 'done'. This can be seen in Algorithm \ref{alg:4}. The robber is also allowed to stay in place with a stay action. Each variant of the problem adds or alters one of the actions. The friendly variant creates a new terminal action which just requires 'caught', while a variable speed robber increases the number of move fluents required.

When running the algorithm, a strong cyclic or strong acyclic solution corresponds to a robber win, while no strong cyclic solution found corresponds to a cop win. In practice, the result will always be a strong cyclic and not a strong acyclic as the game is only a win for the robber if they can avoid capture in all possible states. For a non-cyclic solution to be found, the graph would need to be infinitely large.

Our model uses the PDDL language and the PR2 planner \cite{pr2}. Each problem is given its own domain file to prevent an exponential blow-up of the search space. This exponential blow-up occurs when trying to parameterize the cop's moves. By hard-coding the cops' move to coincide with each node of the graph, we can avoid this issue. For every graph, the problem and domain files are generated programmatically.

Furthermore, to generate the domains for each instance, we also define the problem files for each one. This comes with some customization to control what we solve for. The first addition is the inclusion of a node n0 outside the graph. This node has a unidirectional connection with all possible nodes and cannot be returned to. Beginning with all the cops, each entity makes a move starting at n0 and placing itself on the graph. The cops' nondeterministic behaviour allows all starting configurations to be tested in implicit parallel. The problem files can also be generated with known starting locations. We also allow the user to create instances with variable numbers of cops and robbers. Lastly, digraphs and unconnected nodes are permitted.

\section{Soundness of Encoding}
\label{sec:soundness}

We show that the nondeterministic PDDL model described in Section~\ref{sec:approach} is a correct formalization with respect to the standard rules of the Cops and Robbers problem on a graph, $G$. Our aim is to justify that the state-transition system introduced by our PDDL domain faithfully represents only the legal configurations and moves of the original game. Once this correspondence is established, soundness follows directly from the semantics of nondeterministic planning: every strong cyclic solution returned by the planner corresponds to a valid robber-winning strategy in the original Cops and Robbers problem, and if no strong cyclic solution is found, then the robber has no strategy that avoids capture under all possible cop moves, which means it is $k$-copwin. Thus, soundness reduces to verifying that the PDDL encoding preserves the state representation, action semantics, and termination conditions, turn order, termination, and solution consistency. 

\textbf{State representation.} A state is defined by the truth values of predicates that encode the legal configurations of the Cops and Robbers game. The predicate (at ?x - entity ?y - location) encodes the position of the entity ?x (robber or cops) at its location ?y ($v \in V$), while (edge ?x ?y - location) encodes the adjacency relation of $G$, ensuring all moves occur along graph edges. The predicate (turn ?x - cop) encodes the turn structure: the robber can move iff this is false for all cops, and exactly one cop can move when it's true. The predicate (caught) is true when the robber is occupying the same vertex as any cop. Conversely, (survived) is true when the robber is not occupying the same vertex as any of the cops. Finally, (move) allows for the robber's turn; when this is true, the cop cannot move. For the speedy robber variation, (move0), …, (movek) counts the number of moves the robber can make. The effects of the movement actions modify only these predicates and enforce exactly the legal transitions of the game; every reachable PDDL state corresponds to a legal game position, and every transition corresponds to a legal robber or cop move.

\textbf{Action semantics.} Here, we show that cop and robber moves correspond exactly to the legal game's transitions.

Robber movement: Algorithm \ref{alg:1} has preconditions (at rob ?x), (edge ?x ?y) to ensure the robber must move along an edge from its current vertex, and (forall (?c - cop) (not (turn ?c))) ensures the robber only moves when all cop turns are false. The effects of Algorithm \ref{alg:1} correspond exactly to the game semantics. (not (at rob1 ?x)) and (at rob1 ?y) update the position of the robber from the current vertex ?x to an adjacent vertex ?y. (when (exists (?c - cop) (at ?c ?y)) (caught)) detects capture if a cop is on the vertex that the robber has just moved to. The move counter for the speedy robbers enforces exactly one robber turn-cycle followed by cop turns. The nondeterministic oneof clause models that the robber may or may not survive at a given state to end the game because the game does not end when the robber is caught. This follows the FOND modelling strategy that ensures an agent continues to execute indefinitely \cite{Patrizi_Lipoveztky_Giacomo_Geffner}. Therefore, every robber transition in the PDDL corresponds to a legal robber move in the actual game.

Cop movement: Algorithm \ref{alg:3} requires (turn ?c), ensuring only that the cops can move at this time, and that it is situated at its current node node2. The effect ensures it is no longer this cop's turn and that the robbers now make their move if it was the last cop to move. The oneof clause nondeterministically selects a neighbouring vertex to move to, correctly modelling nondeterministic universal quantification over cop choices. The planner must find a plan that succeeds under all possible outcomes of this nondeterministic choice. This matches the universal quantification in the definition of a robber winning strategy: the robber must succeed against all possible cop responses. Therefore, the cop transitions are sound with respect to nondeterministic branches.

\textbf{Turn alternation.} Robber moves while all (turn ?c) predicates are false. After the robber exhausts its move counter (for the speedy robber, because it gets multiple turns, but the standard robber gets 1 move), the last robber action sets every (turn ?c) to true. Each cop action unsets its own (turn ?c) once it makes a move. When all cop turns are unset, the robber may move again. At every point in any plan, exactly one team (robber or cops) has legal actions, never both, and never neither. This matches the Cops and Robbers turn-based model exactly.

\textbf{Termination.} (caught) becomes true only in states where a cop and a robber occupy the same vertex. (survived) is only true if the robber remains uncaught and the planner's nondeterministic outcome selects continuation. In Algorithm \ref{alg:4}, (done) marks when the game finishes iff (survived) is true and not caught is true, then the game terminates for a given $k$. Thus, the planner may terminate only in states where the robber is alive (survived). If the planner cannot avoid reaching a state with (caught) at some point, then no strong or strong cyclic solution exists. This is exactly the semantics of a robber winning strategy: avoid capture indefinitely. 

\textbf{Correspondence of strong cyclic solution and robber-win strategy.} A strong cyclic solution exists in the PDDL encoding iff the robber has a winning strategy in the Cops and Robbers game on $G$. 

If the planner finds a strong cyclic plan, the PR2 interprets nondeterministic branching as all possible cop choices and evaluates the plan under universal quantification. A strong cyclic solution guarantees that the robber never reaches (caught) along any path. Therefore, the robber has a strategy that defeats all cop behaviours, which implies that the robber wins. Note that there can never be a strong acyclic solution, since these types of policies are finite.

If the robber has a winning strategy, then because the model is sound and complete with respect to legal transitions, every legal robber move is available, and every possible cop move appears as a nondeterministic successor. Therefore, the robber's winning strategy can be encoded as a strong cyclic solution of the planner. Hence, the planner will find one, assuming the underlying planner is sound and complete. This shows the equivalence and soundness.

\textbf{Determining the cop number.} The model enumerates all initial cop placements via nondeterminism. A strong cyclic solution fails exactly when capture is unavoidable under all branches. Running the planner with $k$-cops and observing a solution found implies the robber has evaded capture and the cop number greater than $k$.
If no solution is found, then the robber has been caught and the cop number is $\leq k$. This is sound for determining the cop number of a graph.

Therefore, the overall approach is sound because every PDDL transition corresponds to a legal game transition, every strong cyclic solution corresponds to a robber winning strategy, and the absence of such a plan corresponds to guaranteed cop victory.

\section{Evaluation}
\label{sec:evaluation}

All experiments were conducted on a Dell XPS 17 9720. A 12th Gen Intel(R) Core(TM) i7-12700H was used with 32 GB of available RAM.

An evaluation of the Cops and Robbers problem with planning will be divided into two sections. The first section will apply the technique to classes of problems to empirically evaluate the planner's performance, as well as demonstrate the tool's usefulness for property testing and generation. The second part will look at individual problems. This section will be devoted to interpreting the policy output and determining the algorithm's limits.

\subsection{Property Testing}

Many open questions in Cops and Robbers concern whether certain graph classes obey certain properties. For example, genus can imply bounded cop number \cite{quilliot1985short}. We can evaluate questions of this nature by exhaustively exploring all configurations of the Cops and Robbers problem. If a solution to the FOND problem exists, then the number of cops is insufficient, since the robber can always escape. Likewise, if no solution exists, the robber is unable to generate a plan to avoid capture, and therefore, the minimum number of cops is known.

We evaluated our algorithm on 2 classes of graphs: 7-vertex-connected graphs and 6-vertex Eulerian digraphs. Eulerian digraphs are graphs with directed edges where every node has even in-degree and even out-degree. The exhaustive list of graphs comes from a database constructed by McKay \cite{McKay}. These classes were chosen based on their size and specific properties. Both classes are connected, meaning every node is reachable from every other node. Without this property, we could simply initialize Cops and Robbers on each connected component simultaneously, and play subsequent turns only in the component containing the robber. The Eulerian graph was also chosen to test digraphs within our methodology. Backtracking is not allowed on these graphs. Finally, despite no reflexive edges we do allow for the players to remain on the same vertex.

\begin{table*}[t]
  \centering
  \begin{tabular}{|c|c|c|c|c|c|c|c|}
 \hline
 \textbf{Variation} & \textbf{Total Time} & \textbf{Robber Wins} & \textbf{Weak Searches} & \textbf{Solution Size} & \textbf{Rounds} & \textbf{FSAPS} & \textbf{Poison Count}\\
 \hline
 Standard & 37m39.610s & 450/853 & 361.17 & 54.21 & 10.09 & 197.42 & 158.48\\
 \hline
 Friendly & 33m51.066s & 853/853 & 13 & 10 & 1 & 0 & 0\\
 \hline
 2 Moves & 45m18.327s & 503/853 & 618.94 & 66.59 & 14.09 & 85.23 & 174.83\\
 \hline
 3 Moves & 48m7.595s & 528/835 & 998.47 & 84.13 & 16.61 & 96.01 & 212.66\\
 \hline
 2 Cops & 115m30.003s & 0/835 & 454.44 & 228.56 & 5.70 & 4139.79 & 403.86\\
 \hline
\end{tabular}
\caption{Results over all 7 vertex-connected graphs}
\label{tab:prop1}
\end{table*}

\begin{table*}[t]
  \centering
  \begin{tabular}{|c|c|c|c|c|c|c|c|}
 \hline
 \textbf{Variation} & \textbf{Total Time} & \textbf{Robber Wins} & \textbf{Weak Searches} & \textbf{Solution Size} & \textbf{Rounds} & \textbf{FSAPS} & \textbf{Poison Count}\\
 \hline
 Standard & 89m12.562s & 86/2162 & 71.82 & 32.59 & 4.09 & 397.92 & 45.95\\
 \hline
 Friendly & 92m49.273s & 2162/2162 & 23.68 & 10.95 & 1.35 & 3.69 & 6.49\\
 \hline
 2 Moves & 108m47.285s & 91/2162 & 303.78 & 48.52 & 10.64 & 85.79 & 160.13\\
 \hline
 3 Moves & 144m34.607s & 115/2162 & 849.92 & 64.02 & 22.73 & 138.47 & 329.89\\
 \hline
 2 Cops & 1075m48.238s & 0/2162 & 527.64 & 278.06 & 4.76 & 12952.74 & 460.23\\
 \hline
\end{tabular}
\caption{Results over all 6 vertex Eulerian Digraphs}
\label{tab:prop2}
\end{table*}

Each graph class is explored with some variations. In all variations, each entity starts outside the map and chooses a location, with the robber choosing last. This allows every variation of the game to be considered, as all initial cop configurations will be generated. It is standard for the robber to be allowed to choose their initial position in many cases. The standard variation involves the robber having one move and avoiding the cops. The friendly variation incentivizes the robber to be caught. Each additional move is an extra move a robber may do. Lastly, we increase the number of cops to determine the cop number. That is, we rerun multiple problem instances with gradually increasing number of cops to identify the minimal $k$ that guarantees capture. The results are presented in Tables \ref{tab:prop1} and \ref{tab:prop2}.

Each run measures the following statistics. The first is the time required to run all graphs in the class exhaustively. The next column keeps track of the number of robber-win graphs. This is the number of graphs in which a robber can win by avoiding the cop indefinitely. The robber can always win in the friendly variation, as there is a finite number of states that can be reached. Every state can reach some state in which the robber wins, so all states will eventually lead to a win. The next column shows the average number of weak searches called. This is a measure of how many fast downward planner calls are needed. As the number of moves increases, so too does the number of weak searches on average. Solution size is the average number of states in the controller. This can represent how complex the policies are. FOND problems have no definition of optimality other than one policy dominating another, so this gives some insight into how our policy is, but it cannot be used to say one area is better than another definitively. The next column shows the average number of rounds computed. This, much like the total time, is another measure of computation. A round is defined as restarting the search with a new incumbent solution, using the knowledge learned from previous searches. The more rounds that need to be computed, the more computation is done. The last two columns are forbidden state-action pairs (FSAPs) and poison count. The former is the number of pairs of states and actions that the algorithm is penalized for choosing. The latter is the number of states whose descendants are poisoned. The poisoning process disincentivizes searching down bad paths.

Some interesting patterns emerge when examining the different variations. One of the most significant outliers is the effect of adding an additional cop. One additional cop more than tripled the runtime for the standard version in the first trial. It significantly increased the search space while requiring far fewer searches. This is in part because of the significant number of FSAPs. When adding a cop to the 6-node digraphs, the runtime increased by a factor of 10. 

Compared with increasing the number of cops, increasing the number of moves had a far lesser impact on the computational burden. It added an additional standard deviation's worth of weak searches, which makes intuitive sense. Interestingly, the number of FSAPs decreased. Our guess is that you can enter and leave a forbidden zone without incurring losses, with multiple moves.

Lastly, increasing the number of cops per version allows us to calculate the cop count for the graphs. 403 connected 7-vertex graphs have a cop number of 1, and the rest have a cop number of 2. The digraph results also reveal that the digraph has a maximum cop number of 2.

\subsection{Individual Problems}

Next, we narrow the focus to individual problems to more thoroughly examine their output. By focusing on specific problems, we can evaluate policies directly in terms of human intuition and interpretability, and note areas where our algorithm may be weak. Here, we would like to highlight four different problems that each offer a unique insight into how the algorithm generates policies. This will allow better utilization of the policies in practice. The first problem is a single line in which the cop must chase the robber to the end. This is important for understanding the robber's survivability and how to compute friendly robbers. The second is a square graph. This again offers insight into the friendly robber and whether or not the robber can take advantage of hiding on a square. The third is a tree. This problem demonstrates the issue with backtracking and tests the limits of survivability. The fourth is a tree with a cyclic ring that can only be reached through backtracking. This tests whether a planner can form policies that backtrack if they lead to eventual success. Each problem discussed will be very small so that the policy may be interpretable.

\begin{figure}[t]
\includegraphics[width=6cm]{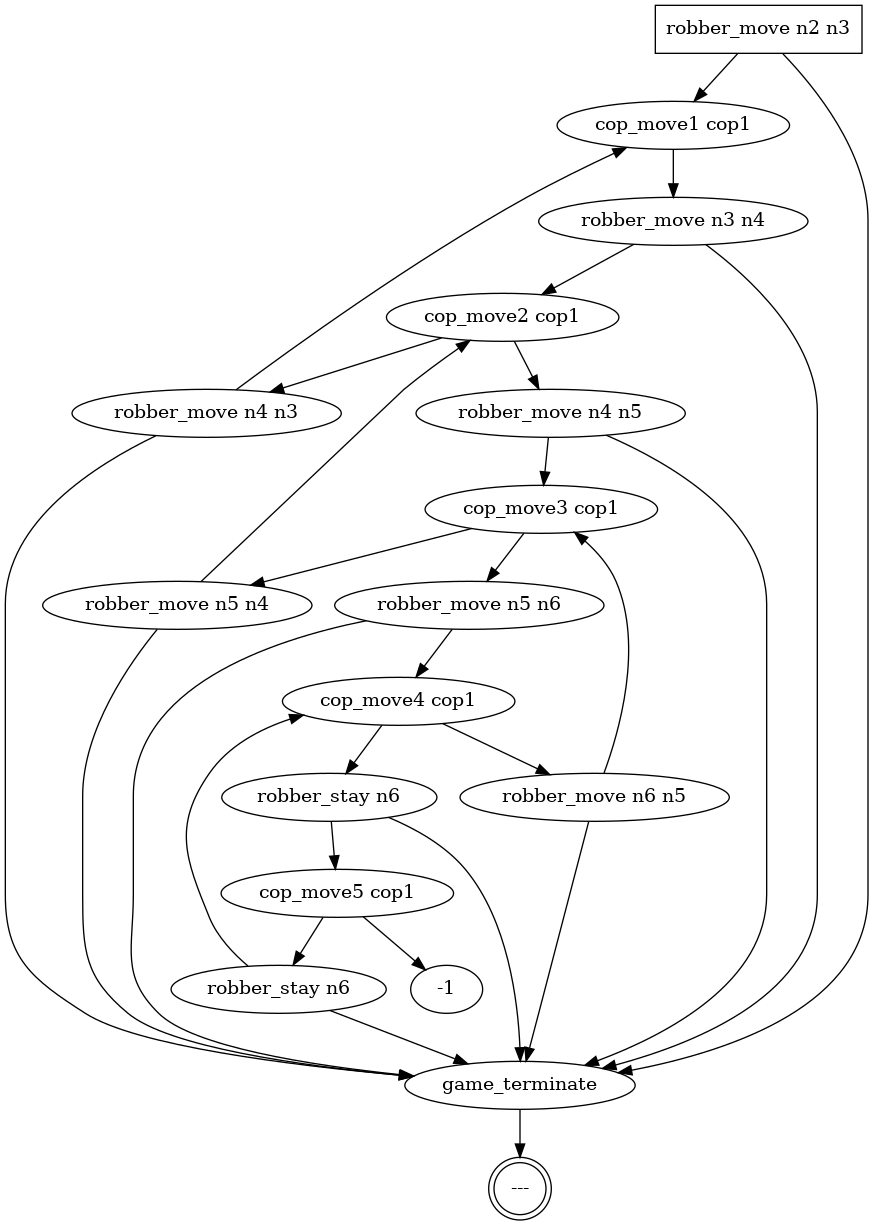}
\centering
\caption{Policy to avoid capture on a straight line}
\label{fig:linechase}
\end{figure}

The first problem to examine is the line chase problem. This problem starts with 6 nodes. The cop begins on the first node, and the robber begins on the second. The policy is shown in Figure \ref{fig:linechase}. 

The policy begins with the action in the square box. From there, one of two results can happen. Either the game ends as the robber survives, leading to the terminal node, or the cop must now make a move. After the cop moves, the robber may choose based on the cop's move. If a node containing -1 is reached, that means that the robber was caught and no strong cyclic solution exists. 

The robber begins by moving away from the cop and down the line. The cop will then chase the robber down. The robber moves away from the cop again, and then a non-deterministic action is taken: the cop either moves towards the robber or back to the beginning. If the cop moves back to the beginning, then the robber will move back towards the cop. The planner always tries to return to known states to minimize the solution space. 

One notable aspect of this problem is the use of the wait action. The robber correctly stays at the final node unless it is safe to move 2 actions toward the cop. This ensures that it survives until the cop reaches the final node. This is the optimal behaviour for surviving as long as possible, and what we would expect a human to do. The plan would also have been valid had instead returned to a previous node, since the cop would have moved back to a known state.

This problem has unique considerations in the case of the friendly robber. Since the planner tends to return to known locations, the robber does not take the shortest path to the cop. Instead, they repeatedly return to known positions until the cop makes novel choices. To avoid this scenario and generate the shortest plans, the additional parameter "--localize-enabled 0" may be added to the PR2 planner \cite{pr2}. This disables the check for known states, and instead, the policy is returned where the robber moves down the line with each move.

The second notable plan is the square. This is the shortest possible cyclic plan in which the robber has a winning strategy. A triangle is infeasible as it is a fully connected graph.

\begin{figure}[t]
\includegraphics[width=\linewidth]{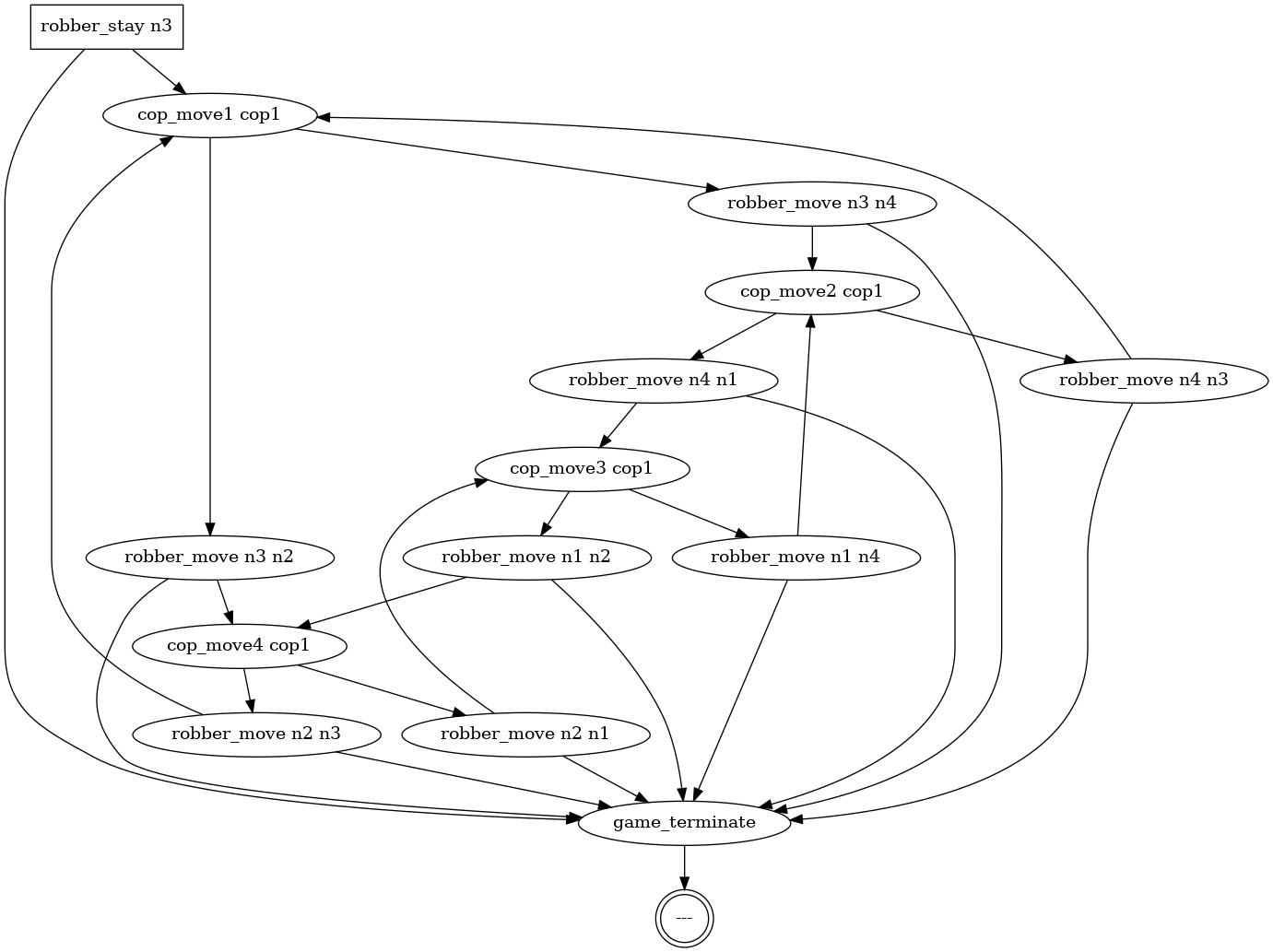}
\centering
\caption{Policy to avoid capture on a square}
\label{fig:square}
\end{figure}

This graph shows that the robber can wait in its initial state before acting on the cop's action. It will always stay in the opposite corner, which is the intended behaviour. The full policy is shown in Figure \ref{fig:square}. This problem also gives interesting behaviour when the robber is given multiple moves. If given 2 moves, the robber will always wait on its second move, since taking a move would put it in a position to be captured the following turn. If given an odd number of moves, such as three, the robber will be instructed to make 2 moves which cancel each other out. It will move to a node only to move back. This pattern seems to be repeated in most multiple-move graphs. The planner's additional goal of returning to a goal state makes the addition of repeated moves often negligible, as the robber is incentivized to waste them, unless necessary.

Adding a cycle also affects how the friendly variation works. We cannot use the additional parameter ``--localize-enabled 0'' as before. In this scenario, it is always possible for the cop to move away from the robber by moving to the node opposite on the cycle with respect to the robber. Thus, we have to rely on returning to additional states. Unrolling the complete paths can result in paths that are infinitely long, which is beyond the scope of this planner.

The third problem worth inspecting is a branching tree problem. This problem tests the robber's ability to make long-term decisions even if the outcome is the same. That is, can the robber choose the longest branch even if, in all cases, it will eventually be a cop win? The answer is yes. The policy can be seen in Figure \ref{fig:branch}. The policies for the other problems can be found in the Appendix.

\begin{figure}
\includegraphics[width = 9cm, height=6cm]{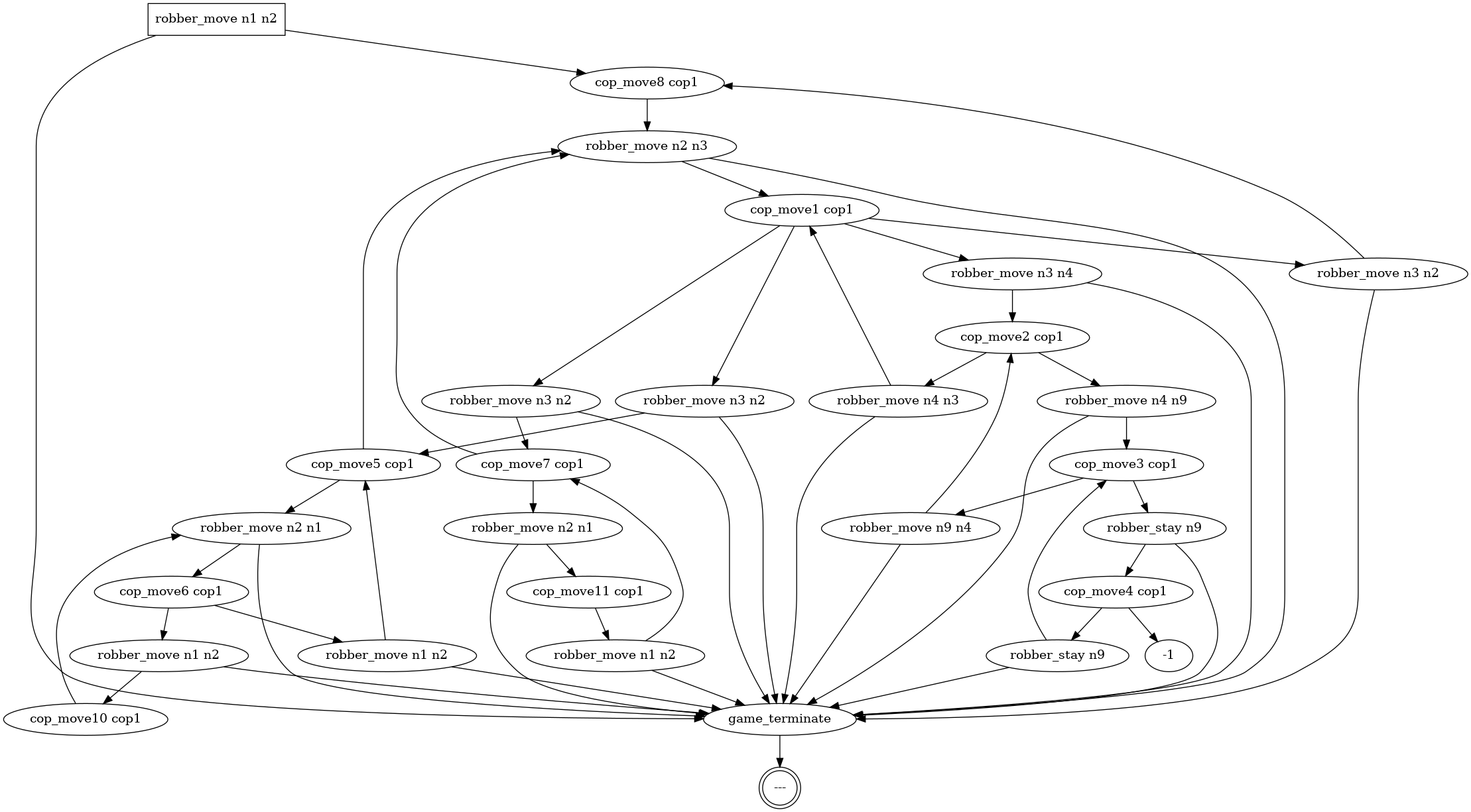}
\centering
\caption{Policy to avoid capture on a Tree}
\label{fig:branch}
\end{figure}

In this problem, the robber begins at the root of the tree, and the cop begins adjacent to the robber on the shortest branch. From this scenario, the robber is given the option of which branch to hide down. The robber picks the longest branch and evades capture for as long as possible. This is again notable as all choices lead to valid plans, yet if we want to create plans that lead to longer games, this is the behaviour we would hope to see.

The 4th problem changes the prior problem in 2 different ways. Both involve making the branch the cop starts on the longest. The first version simply makes it the longest by adding nodes to extend the line. The second version adds a loop of size 4 to the branch. This not only makes it the longest but also ensures that if the robber were to reach the cycle, they have a guaranteed win.

Unfortunately, the domain and problem make the robber incapable of performing the backtracking required to descend this path. The way to access the longest branch would be to wait for the cop to descend a branch the robber did not access, then backtrack to the longer one. In both cases, the planner failed to discover this policy. 

This failure to generate the policy also extended to parameter tuning. Increasing the search depth and the number of trials, as well as disabling known-state returns, failed to cause the planner to backtrack. This shows that, in its current state, the planner will not survive for as long as possible.

This, however, does not suggest that the planner cannot find solutions. In all backtracking cases, there is a possibility of guaranteed failure if a cop immediately descends the chosen branch of the robber. As a result, there is no strong cyclic solution. We were unable to develop a problem that requires backtracking but also has a strong cyclic solution. Therefore, it is inconclusive if any solutions may be missed due to this aspect of the program.

\section{Summary}
\label{sec:summary}

Cops and Robbers problems are a class of problems of significant interest in graph theory and mathematics \cite{Nowakowski1983, scott2011bound, chudnovsky2024cops}. This paper presents 3 main results. The first is FOND encoding of the Cops and Robbers problem. The addition of the non-deterministic cop and variable starts allows for an exhaustive exploration of a given graph, answering the core problem of ``is this graph $k$-cop win?''.
The second contribution is a proof-of-concept and exploration of property testing for classes of problems. We design and generate domains and problems based on graph groups and exhaustively explore the search space. The planner, alongside our implementation, can run on classes of instances and test hypotheses. It can also be used to determine upper and lower bounds. Our implementation is efficient for smaller instances when exploring properties and can be used on large instances for individual problems. The number of cops plays a significant role in how large the problem can be.
The final contribution is regarding policy interpretation. The policies can be extracted and understood intuitively. They follow the rules one would expect of a robber. Some qualities of a plan, such as backtracking, do not quite work in a human-interpretable way, while others, such as waiting out a cop, do appear to work as intended.

Ultimately, we have demonstrated how systematic planning and modelling of a mathematical problem setting can be used to explore properties of interest to the mathematics community. Our approach to the Cops and Robbers problem offers a unique perspective and enables researchers to prove properties of the problem.

Moving forward, we hope to explore additional variants of the Cops and Robbers problem. In this paper, we cover the Friendly Robber and the Variable-Speed Robber. While some variants, such as the drunk robber, would be superfluous with our implementation, most others should be compatible with our framework without substantial adjustment, such as the infinite speed robber \cite{frieze2010variationscopsrobbers}. Further optimizations to the plan generation would also allow for the exploration of larger search spaces. This is particularly important for allowing properties involving many moving entities to be studied. Improving the implementation to support a broader range of problems would yield much stronger use cases for the model. The current model is only tested on smaller graphs, and while graphs with more nodes are feasible, the problem's time complexity is exponential, making it difficult to solve the for arbitrary graphs and we are trying to identify subsets of graphs with properties that are theoretically interesting. We have shown that there are some cases in which human intuition does not match the plan, such as in the tree search with backtracking. With further exploration, we believe this can be improved.

\bibliography{aaai25}

@article{Frieze2012,
   author = {Alan Frieze and Michael Krivelevich and Po Shen Loh},
   doi = {10.1002/jgt.20591},
   issn = {10970118},
   issue = {4},
   journal = {Journal of Graph Theory},
   pages = {383-402},
   publisher = {Wiley-Liss Inc.},
   title = {Variations on cops and robbers},
   volume = {69},
   year = {2012},
}

@article{Patrizi_Lipoveztky_Giacomo_Geffner, journal = {IJCAI}, title={Computing Infinite Plans for LTL Goals Using a Classical Planner}, abstractNote={Classical planning has been notably successful in synthesizing ﬁnite plans to achieve states where propositional goals hold. In the last few years, classical planning has also been extended to incorporate temporally extended goals, expressed in temporal logics such as LTL, to impose restrictions on the state sequences generated by ﬁnite plans. In this work, we take the next step and consider the computation of inﬁnite plans for achieving arbitrary LTL goals. We show that inﬁnite plans can also be obtained efﬁciently by calling a classical planner once over a classical planning encoding that represents and extends the composition of the planning domain and the Bu¨chi automaton representing the goal. This compilation scheme has been implemented and a number of experiments are reported.}, author={Patrizi, Fabio and Lipoveztky, Nir and Giacomo, Giuseppe De and Geffner, Hector}, language={en}, year={2011} }

@misc{Nowakowski1983,
   author = {Richard Nowakowski and Peter Winkler},
   title = {VERTEX-TO-VERTEX PURSUIT IN A GRAPH},
year = {1983}
}

@misc{nussbaum,
   author = {Doron Nussbaum and Alper Yörükçü},
   title = {Moving Target Search with Subgoal Graphs*},
   url = {www.aaai.org},
    year={2015},
}

@article{guibas1999visibility,
  title={A visibility-based pursuit-evasion problem},
  author={Guibas, Leonidas J and Latombe, Jean-Claude and LaValle, Steven M and Lin, David and Motwani, Rajeev},
  journal={International Journal of Computational Geometry \& Applications},
  volume={9},
  number={04n05},
  pages={471--493},
  year={1999},
  publisher={World Scientific}
}

@INPROCEEDINGS{9341031,

  author={Stiffler, Nicholas M. and O’Kane, Jason M.},

  booktitle={2020 IEEE/RSJ International Conference on Intelligent Robots and Systems (IROS)}, 

  title={Planning for robust visibility-based pursuit-evasion}, 

  year={2020},

  volume={},

  number={},

  pages={6641-6648},

  doi={10.1109/IROS45743.2020.9341031}}

@article{QUILLIOT198655,
title = {Some Results about Pursuit Games on Metric Spaces Obtained Through Graph Theory Techniques},
journal = {European Journal of Combinatorics},
volume = {7},
number = {1},
pages = {55-66},
year = {1986},
issn = {0195-6698},
doi = {https://doi.org/10.1016/S0195-6698(86)80017-8},
url = {https://www.sciencedirect.com/science/article/pii/S0195669886800178},
author = {A. Quilliot}
}

@article{quilliot1985short,
  title={A short note about pursuit games played on a graph with a given genus},
  author={Quilliot, Alain},
  journal={Journal of combinatorial theory, Series B},
  volume={38},
  number={1},
  pages={89--92},
  year={1985},
  publisher={Elsevier}
}

@article{komarov2013,
author = {Komarov, Natasha and Winkler, Peter},
year = {2013},
month = {05},
pages = {},
title = {Capturing the Drunk Robber on a Graph},
volume = {21},
journal = {Electronic Journal of Combinatorics},
doi = {10.37236/3398}
}

@article{Harris2020,
   author = {Pamela Harris and Erik Insko and Alicia Prieto-Langarica and Rade Stoisavljevic and Shaun Sullivan},
   month = {4},
   title = {Tipsy cop and drunken robber: a variant of the cop and robber game on graphs},
   url = {http://arxiv.org/abs/2004.00606},
   year = {2020},
}

@article{Kehagias2013,
   author = {Athanasios Kehagias and Dieter Mitsche and Paweł Prałat},
   doi = {10.1016/j.tcs.2013.01.032},
   issn = {03043975},
   journal = {Theoretical Computer Science},
   month = {4},
   pages = {100-120},
   title = {Cops and invisible robbers: The cost of drunkenness},
   volume = {481},
   year = {2013},
}

@misc{macindoe2023,
   author = {Owen Macindoe and Leslie Pack Kaelbling and Tomás Lozano-Pérez},
   title = {POMCoP: Belief Space Planning for Sidekicks in Cooperative Games},
   url = {www.aaai.org},
year = {2023}
}

@misc{McKay, year= {2023}, note = {Accessed: 2023-11-12}, title = {Combinatorial Data}, howpublished = {\url{http://users.cecs.anu.edu.au/~bdm/data/}}, journal={Combinatorial Data}, author={McKay, Brendan}, urldate = {2023-11-19}}

@article{pr2, title={PRP Rebooted: Advancing the State of the Art in FOND Planning}, url={http://arxiv.org/abs/2312.11675}, abstractNote={Fully Observable Non-Deterministic (FOND) planning is a variant of classical symbolic planning in which actions are nondeterministic, with an action’s outcome known only upon execution. It is a popular planning paradigm with applications ranging from robot planning to dialogue-agent design and reactive synthesis. Over the last 20 years, a number of approaches to FOND planning have emerged. In this work, we establish a new state of the art, following in the footsteps of some of the most powerful FOND planners to date. Our planner, PR2, decisively outperforms the four leading FOND planners, at times by a large margin, in 17 of 18 domains that represent a comprehensive benchmark suite. Ablation studies demonstrate the impact of various techniques we introduce, with the largest improvement coming from our novel FONDaware heuristic.}, note={arXiv:2312.11675 [cs]}, number={arXiv:2312.11675}, publisher={arXiv}, author={Muise, Christian and McIlraith, Sheila A. and Beck, J. Christopher}, year={2023}, month=dec, language={en} }

@article{pralat2016meyniel,
  title={Meyniel's conjecture holds for random graphs},
  author={Pra{\l}at, Pawe{\l} and Wormald, Nicholas},
  journal={Random Structures \& Algorithms},
  volume={48},
  number={2},
  pages={396--421},
  year={2016},
  publisher={Wiley Online Library}
}

@article{chudnovsky2024cops,
  title={Cops and robbers on-free graphs},
  author={Chudnovsky, Maria and Norin, Sergey and Seymour, Paul D and Turcotte, J{\'e}r{\'e}mie},
  journal={SIAM Journal on Discrete Mathematics},
  volume={38},
  number={1},
  pages={845--856},
  year={2024},
  publisher={SIAM}
}

@inproceedings{enright2023cops,
  title={Cops and robbers on multi-layer graphs},
  author={Enright, Jessica and Meeks, Kitty and Pettersson, William and Sylvester, John},
  booktitle={International Workshop on Graph-Theoretic Concepts in Computer Science},
  pages={319--333},
  year={2023},
  organization={Springer}
}

@article{kinnersley2015cops,
  title={Cops and robbers is exptime-complete},
  author={Kinnersley, William B},
  journal={Journal of Combinatorial Theory, Series B},
  volume={111},
  pages={201--220},
  year={2015},
  publisher={Elsevier}
}

@article{scott2011bound,
  title={A bound for the cops and robbers problem},
  author={Scott, Alex and Sudakov, Benny},
  journal={SIAM Journal on Discrete Mathematics},
  volume={25},
  number={3},
  pages={1438--1442},
  year={2011},
  publisher={SIAM}
}

@article{lu2012meyniel,
  title={On Meyniel's conjecture of the cop number},
  author={Lu, Linyuan and Peng, Xing},
  journal={Journal of Graph Theory},
  volume={71},
  number={2},
  pages={192--205},
  year={2012},
  publisher={Wiley Online Library}
}

@article{frankl1987cops,
  title={Cops and robbers in graphs with large girth and Cayley graphs},
  author={Frankl, Peter},
  journal={Discrete Applied Mathematics},
  volume={17},
  number={3},
  pages={301--305},
  year={1987},
  publisher={Elsevier}
}

@article{drunkrobber2012,
title = {Some remarks on cops and drunk robbers},
journal = {Theoretical Computer Science},
volume = {463},
pages = {133-147},
year = {2012},
note = {Special Issue on Theory and Applications of Graph Searching Problems},
issn = {0304-3975},
doi = {https://doi.org/10.1016/j.tcs.2012.08.016},
url = {https://www.sciencedirect.com/science/article/pii/S030439751200789X},
author = {Athanasios Kehagias and Paweł Prałat}
}

@misc{frieze2010variationscopsrobbers,
      title={Variations on Cops and Robbers}, 
      author={Alan Frieze and Michael Krivelevich and Po-Shen Loh},
      year={2010},
      eprint={1004.2482},
      archivePrefix={arXiv},
      primaryClass={math.CO},
      url={https://arxiv.org/abs/1004.2482}, 
}

@article{CLARKE2025394,
title = {Time-delayed Cops and Robbers},
journal = {Discrete Applied Mathematics},
volume = {360},
pages = {394-405},
year = {2025},
issn = {0166-218X},
doi = {https://doi.org/10.1016/j.dam.2024.09.030},
url = {https://www.sciencedirect.com/science/article/pii/S0166218X24004207},
author = {Nancy E. Clarke and Danielle Cox and Melissa A. Huggan and Svenja Huntemann and Trent G. Marbach}
}

@misc{bonato2017characterizationsalgorithmsgeneralizedcops,
      title={Characterizations and algorithms for generalized Cops and Robbers games}, 
      author={Anthony Bonato and Gary MacGillivray},
      year={2017},
      eprint={1704.05655},
      archivePrefix={arXiv},
      primaryClass={math.CO},
      url={https://arxiv.org/abs/1704.05655}, 
}

@article{kenter2025improved,
  title={Improved bounds on the cop number when forbidding a minor},
  author={Kenter, Franklin and Meger, Erin and Turcotte, J{\'e}r{\'e}mie},
  journal={Journal of Graph Theory},
  volume={108},
  number={3},
  pages={620--646},
  year={2025},
  publisher={Wiley Online Library}
}

\section{Appendix}

\begin{figure*}[!htbp] 
\includegraphics[width=15cm]{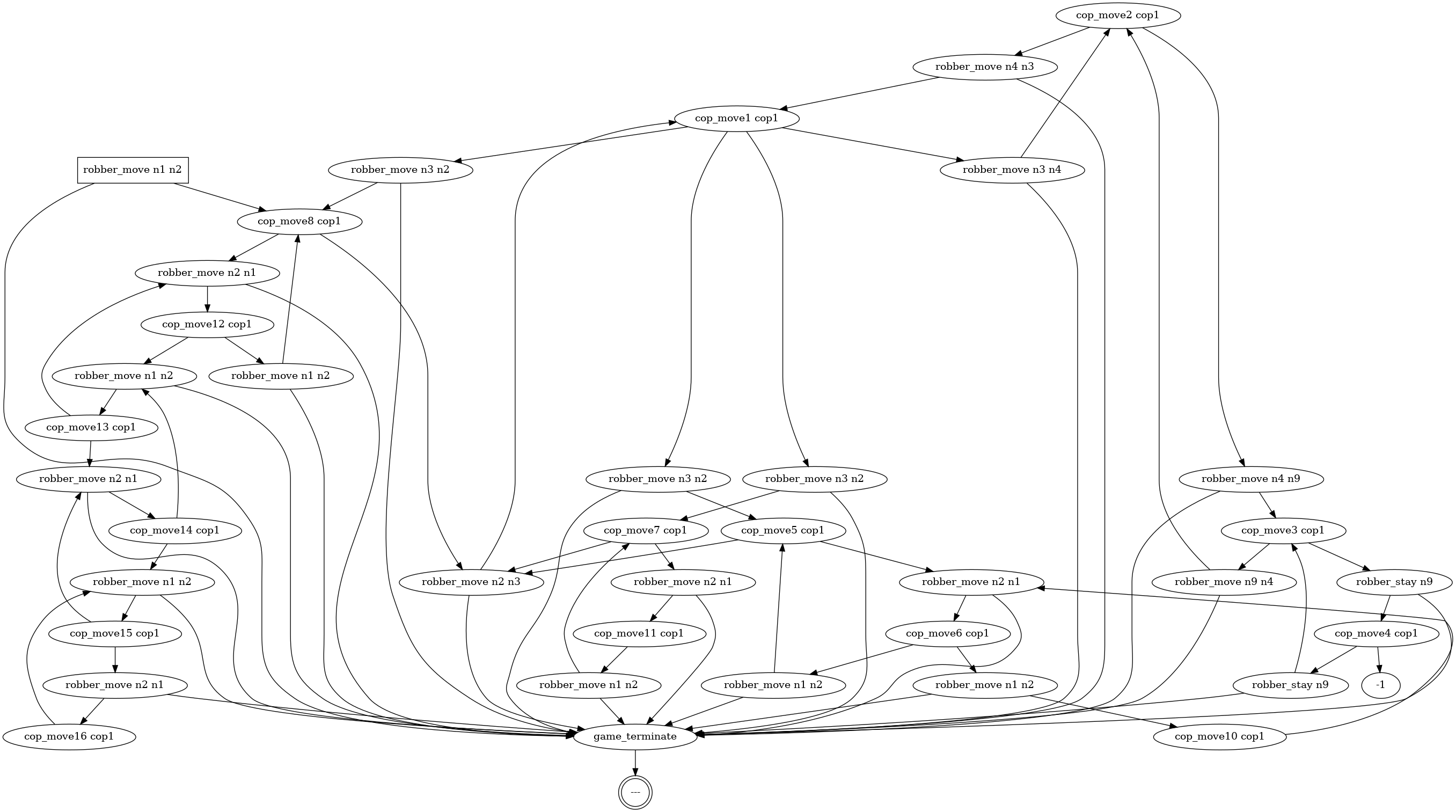}
\centering
\caption{Policy to avoid capture on a tree with backtracking}
\label{fig:tree2}
\end{figure*}

\begin{figure*}[!htbp] 
\includegraphics[width=15cm]{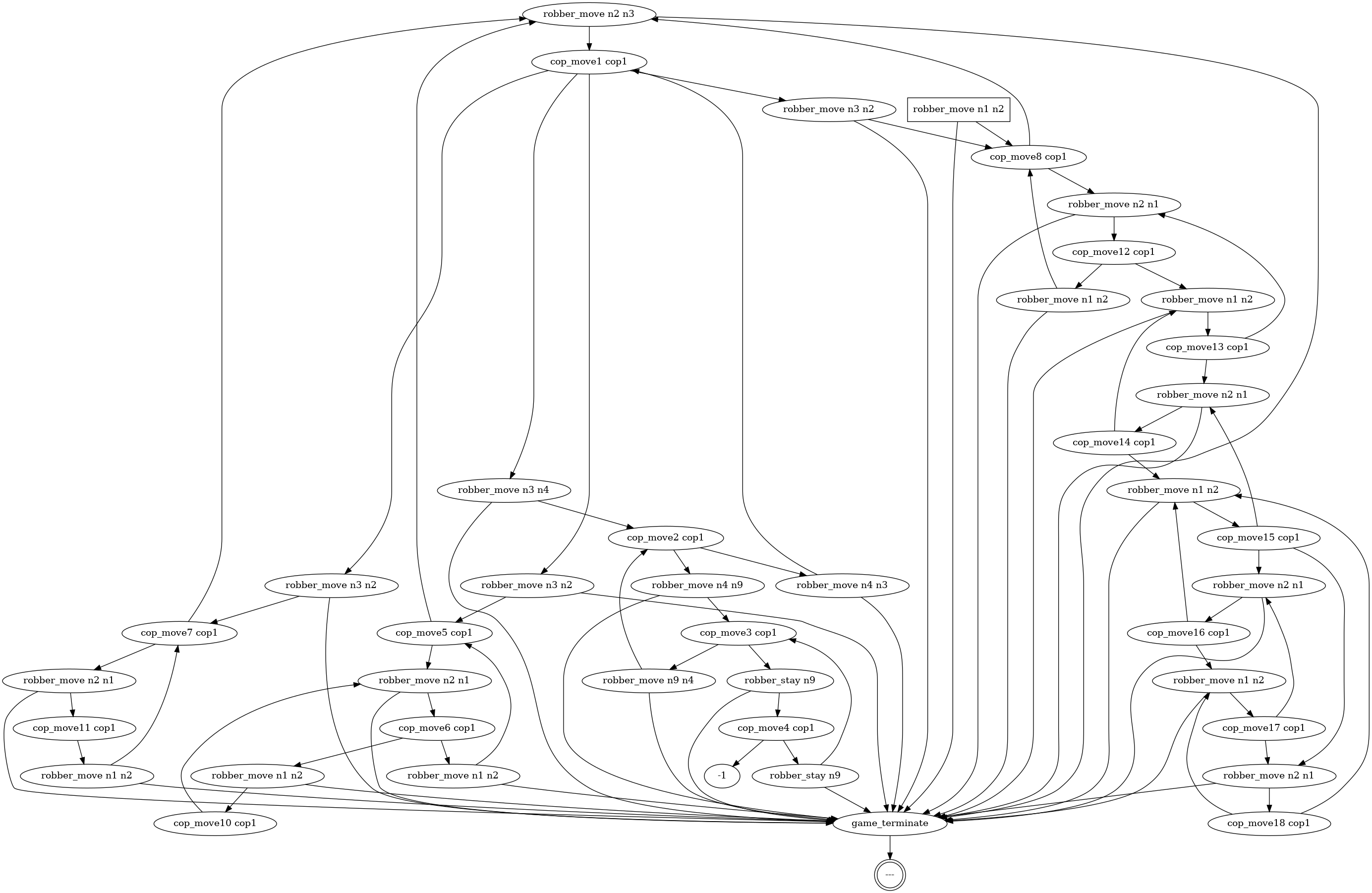}
\centering
\caption{Policy to avoid capture on a tree with a cycle}
\label{fig:stree3}
\end{figure*}

\end{document}